\documentclass[namedreferences]{solarphysics}

\usepackage[hyperref,optionalrh]{spr-sola-addons} 
\usepackage{graphicx}        
\usepackage{color}           

\newcommand{\aap}{    {\it Astron. Astrophys.}}

\newcommand{\apj}{    {\it Astrophys. J.}}
\newcommand{\apjl}{   {\it Astrophys. J. Lett.}}
\newcommand{\apjs}{    {\it Astrophys. J. Suppl.}}

\newcommand{\jgr}{    {\it J. Geophys. Res.}}

\newcommand{\solphys}{{\it Solar Phys.}}
 
\newcommand{\ssr}{    {\it Space Sci. Rev.}}

\begin{document}

\begin{article}

\begin{opening}

\title{Interplanetary Propagation Behavior of the Fast Coronal Mass Ejection from 23 July 2012}

\author{M.~\surname{Temmer}$^{1,2}$\sep  
       N.V. Nitta$^{2}$}
\runningauthor{M.~Temmer and N.V.~Nitta}
\runningtitle{Interplanetary propagation behavior of the CME from 23 July 2012}

   \institute{$^{1}$ Kanzelh\"ohe Observatory/IGAM, Institute of Physics, University of Graz,
Universit\"atsplatz 5, A-8010 Graz, Austria \\  email: \url{manuela.temmer@uni-graz.at} \\
             $^{2}$ Lockheed Martin STAR Labs, Building 252, 3251 Hanover Street, Palo Alto, CA 94304, USA 
          }

\begin{abstract}
The fast coronal mass ejection (CME) from 23 July 2012 raised attention due to its extremely short transit time from Sun to 1~AU of less than 21~h. In-situ data from STEREO-A revealed the arrival of a fast forward shock with a speed of more than 2200~km~s$^{-1}$ followed by a magnetic structure moving with almost 1900~km~s$^{-1}$. We investigate the propagation behavior of the CME shock and magnetic structure with the aim to reproduce the short transit time and high impact speed as derived from in-situ data. We carefully measure the 3D kinematics of the CME using the graduated cylindrical shell model, and obtain a maximum speed of 2580$\pm$280~km~s$^{-1}$ for the CME shock and of 2270$\pm$420~km~s$^{-1}$ for its magnetic structure. Based on the 3D kinematics, the drag-based model (DBM) reproduces the observational data reasonably well. To successfully simulate the CME shock, we find that the ambient flow speed should be of average value close to the slow solar wind speed (450~km~s$^{-1}$), and the initial shock speed at a distance of 30~$R_{\odot}$ should not exceed $\approx$2300~km~s$^{-1}$, otherwise it would arrive much too early at STEREO-A. The model results indicate that an extremely low aerodynamic drag force is exerted on the shock, smaller by one order of magnitude compared to the average. As a consequence, the CME hardly decelerates in interplanetary space and maintains its high initial speed. The low aerodynamic drag can only be reproduced when reducing the density of the ambient solar wind flow, in which the massive CME propagates, to $\rho_{\rm sw}$\,=\,1--2~cm$^{-3}$ at the distance of 1~AU. This result is consistent with the preconditioning of interplanetary space owing to a previous CME.  
 
\end{abstract}
\end{opening}

\section{Introduction}

The coronal mass ejection (CME) from 23 July 2012 was one of the most energetic events ever recorded, having a transit time from Sun to 1~AU of less than 21~h. The event raised a lot of attention due the strong magnetic ejecta and the extremely high impact speed of more than 2200~km~s$^{-1}$ for the shock ahead of it as measured by in-situ instruments aboard STEREO-A/PLASTIC \citep{russell13}. If directed towards Earth, the CME might have been of unusual high geoeffectiveness and scenarios of extreme space weather consequences were proposed \citep{ngwira13,baker13}. \cite{liu14} studied the solar perspective of this event and came to the conclusion that the complex magnetic field and high field strength as measured in-situ is most probably attributable to a CME-CME interaction process. These authors also speculated that the extremely high speed and short propagation time were caused by the increased upstream solar wind speed and decreased solar wind density as well as magnetic field tension, which in turn resulted from the CME associated with an M7.7 flare on 19 July 2012.

The structure of the ambient magnetic field and plasma flow in which CMEs are embedded in may play a critical role in 
controlling their propagation behavior and geoeffectiveness \citep[see \textit{e.g.},][]{vrsnak07,gopal08}. In recent years the stereoscopic view made possible by the \textit{Solar-Terrestrial Relations Observatory} \citep[STEREO;][]{kaiser08} has permitted us to learn much about the evolution of CMEs in the inner heliosphere. Observational results revealed that variations in the ambient solar wind speed, \textit{e.g.}\,by slow and fast solar winds and various interactions, may appreciably change the CME kinematics (including deflection) and deform the CME structure \citep[see \textit{e.g.},][]{rouillard10,temmer11,lugaz12,liu13}. Recent studies pointed out that magnetic erosion by reconnection processes at the front of the magnetic structure of the CME may significantly reduce the strength of geomagnetic storms \citep{ruffenach12,lavraud14}. As such, the preconditioning of interplanetary space has immediate consequences for predicting arrival times and impact speeds of CMEs.

Since the plasma and magnetic field conditions in interplanetary space are not observationally well-known, research on the interaction of CMEs with their environment may be performed by combining the observational data with models. In this study, we investigate the CME from 23 July 2012 in terms of its evolution in the inner heliosphere. The three-dimensional (3D) kinematical profile of the CME is derived up to a distance range of 30~$R_{\odot}$ by applying the graduated cylindrical shell (GCS) model \citep{thernisien06}. Assuming that the aerodynamic drag governs the evolution of the CME at farther distances from the Sun, we use the drag-based model (DBM;\,\cite{vrsnak13}, and references therein) to reconstruct the CME propagation behavior. This yields results that strongly indicate the low density environment in which the CME is propagating in. 

In Section~\ref{obs} we give an overview of the observational data and describe the methods that are used for the analysis. We give the results in Section~\ref{res} and discuss them and draw our conclusions in Sections~\ref{dis} and~\ref{con}.

\section{Observational Data and Methods}\label{obs}

On 23 July 2012 at around 02:08~UT a fast CME is launched from solar active region NOAA~11520 located at W141$^{\circ}$, \textit{i.e.}\,behind the limb from Earth-view. Two filament eruptions may have driven two CMEs, which then 
interacted \citep[see][]{liu14}. However, we cannot clearly separate two individual CME profiles in the coronagraph data. Therefore, for our study we simply refer to the event as two-stage filament eruption resulting in a complex CME, consistent with the overall CME profile from \cite{liu14}. In this respect we note that the derived longitude range of the CME is consistent with the erupting filament located at the eastern edge of the active region. For the event time, the two STEREO spacecraft, STEREO-A (\textit{Ahead}) and STEREO-B (\textit{Behind}), were located, respectively, at the heliographic longitude of W121$^{\circ}$ and E115$^{\circ}$, and heliocentric distance of 0.96~AU  ($\approx$206~$R_{\odot}$) and 1.02~AU ($\approx$219~$R_{\odot}$). STEREO-A views the CME as halo event, STEREO-B close to its eastern limb with the plane-of-sky angle W155$^{\circ}$. We use the disk-integrated flux from the EUVI-A 195\AA~channel as proxy to estimate the soft X-ray emission \citep[see][]{nitta13}. According to this, the associated flare is a long duration event of at most X2.5 GOES class.

The development of the CME up to a distance range of $\approx$30~$R_{\odot}$ is observed from combined EUV and white-light data from STEREO-A and -B as well as the \textit{Solar Heliospheric Observatory} \citep[SoHO;][]{domingo95}. Using contemporaneous imaging data from the coronagraphs aboard those spacecraft enables us to do a 3D reconstruction of the CME evolution by applying the GCS model. The different viewpoints from at least two spacecraft (combination of COR2-A on STEREO-A, COR2-B on STEREO-B, and  LASCO/C2/C3 on SoHO) are digested into the model and by forward fitting the model flux rope to the observed white-light structure we obtain the 3D geometry (width and cross section), propagation direction and deprojected kinematics of the CME. We apply the same model for investigating the shock ahead of the flux rope by putting the half-angle to zero and the ratio aspect close to one. This makes the front part of the GCS model spherical in order to mimic the geometry of a shock front \citep[see][]{thernisien11}. Measurements coming from single spacecraft data (EUVI-B, COR1-B) are deprojected onto the derived 3D results using a multiplication factor. From the deprojected distance-time data we derive the speed- and acceleration time profiles by using the regularization method originally developed by \cite{kontar04} to invert solar X-ray spectra measured by the \textit{Reuven Ramaty High-Energy Solar Spectroscopic Imager} \citep[RHESSI;][]{lin02}. The method was further developed and could be successfully applied to CME kinematical data as described in \cite{temmer10}.

By calibrating white-light images from COR2-B data in units of the mean solar brightness and subtracting a pre-event image we calculate the excess brightness due to the CME. The excess brightness is then converted to the excess mass of the CME under the common assumptions that all of the CME mass is concentrated on the plane-of-sky of the corresponding instrument, and that the CME material consists of 90\%~H and 10\%~He. The brightness of all pixels in the region covering the CME is then summed up and fed into the following relation $m=B_{\rm obs}/B_{\rm e}(\Phi)$, where $B_{\rm obs}$ is the white-light pixel brightness and $B_{\rm e}(\Phi)$ the single electron brightness at the angle $\Phi$ from the plane-of-sky \citep[see][]{billings66,vourlidas00}. 

In-situ data are taken from the STEREO-A/IMPACT-MAG instrument which measures the magnetic field direction and magnitude \citep{luhmann08}. The high fluxes of solar energetic particles associated with this event are the cause of a data gap for the solar wind plasma data by PLASTIC \citep{galvin08} and, hence, are reconstructed from the magnetic field data (5~min resolution). This introduces uncertainties of 1\% for reconstructed speeds below 900~km~s$^{-1}$ and for the peak of the order of 200~km~s$^{-1}$ (A.B. Galvin, private communication). For our study we extract the arrival time and the lower limit of the impact speed of the CME shock and magnetic ejecta as given in \cite{russell13}.

\begin{figure}
\centerline{\includegraphics[width=1\textwidth,clip=]{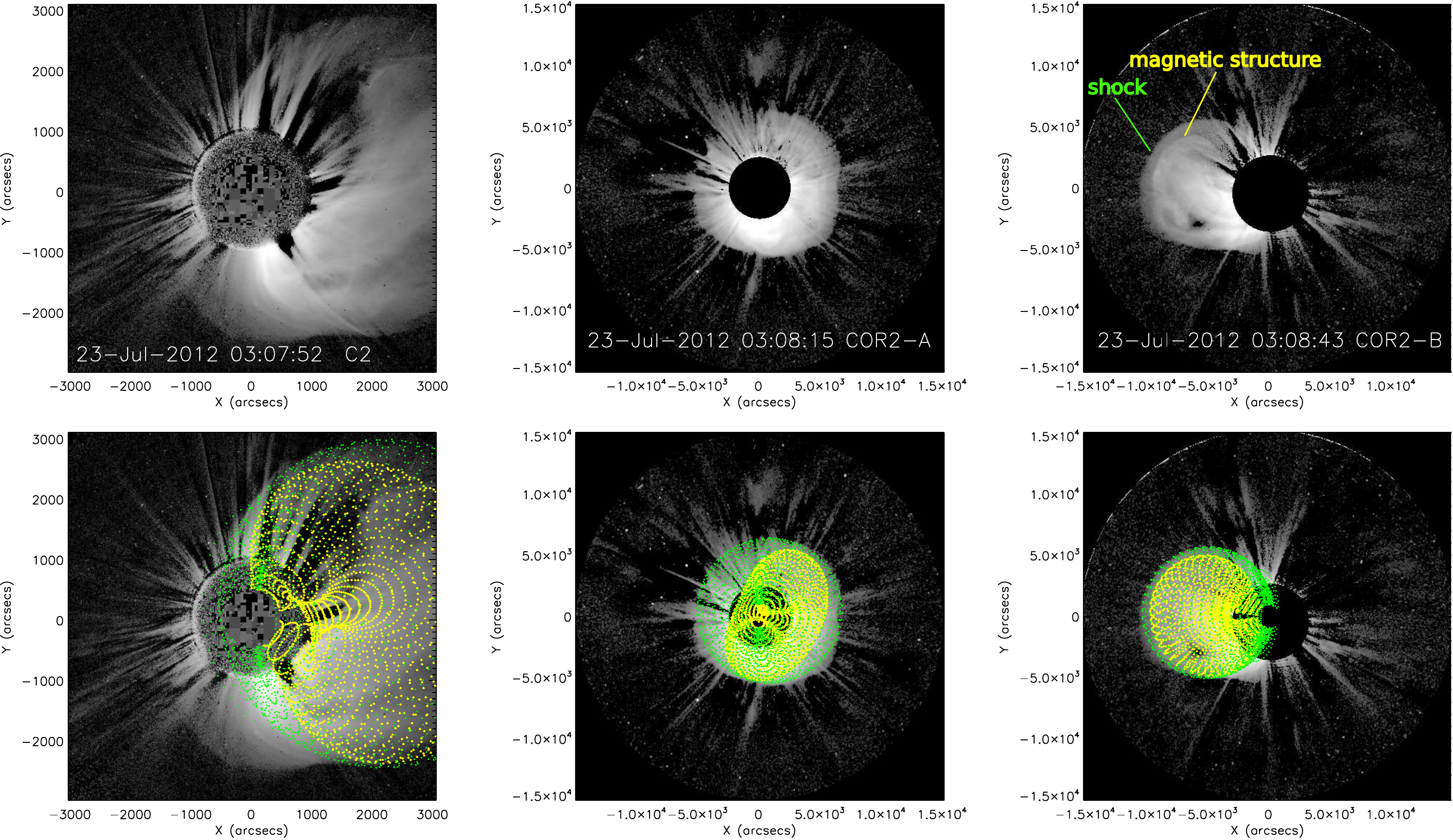}}
 \caption{GCS model results for the reconstructed magnetic structure (yellow mesh) and shock (green mesh) using simultaneous image triplets of COR2-A, COR2-B, and LASCO-C2 at $\approx$03:08~UT.}
	\label{gcs}
\end{figure}

\begin{figure}
\centerline{\includegraphics[width=1\textwidth,clip=]{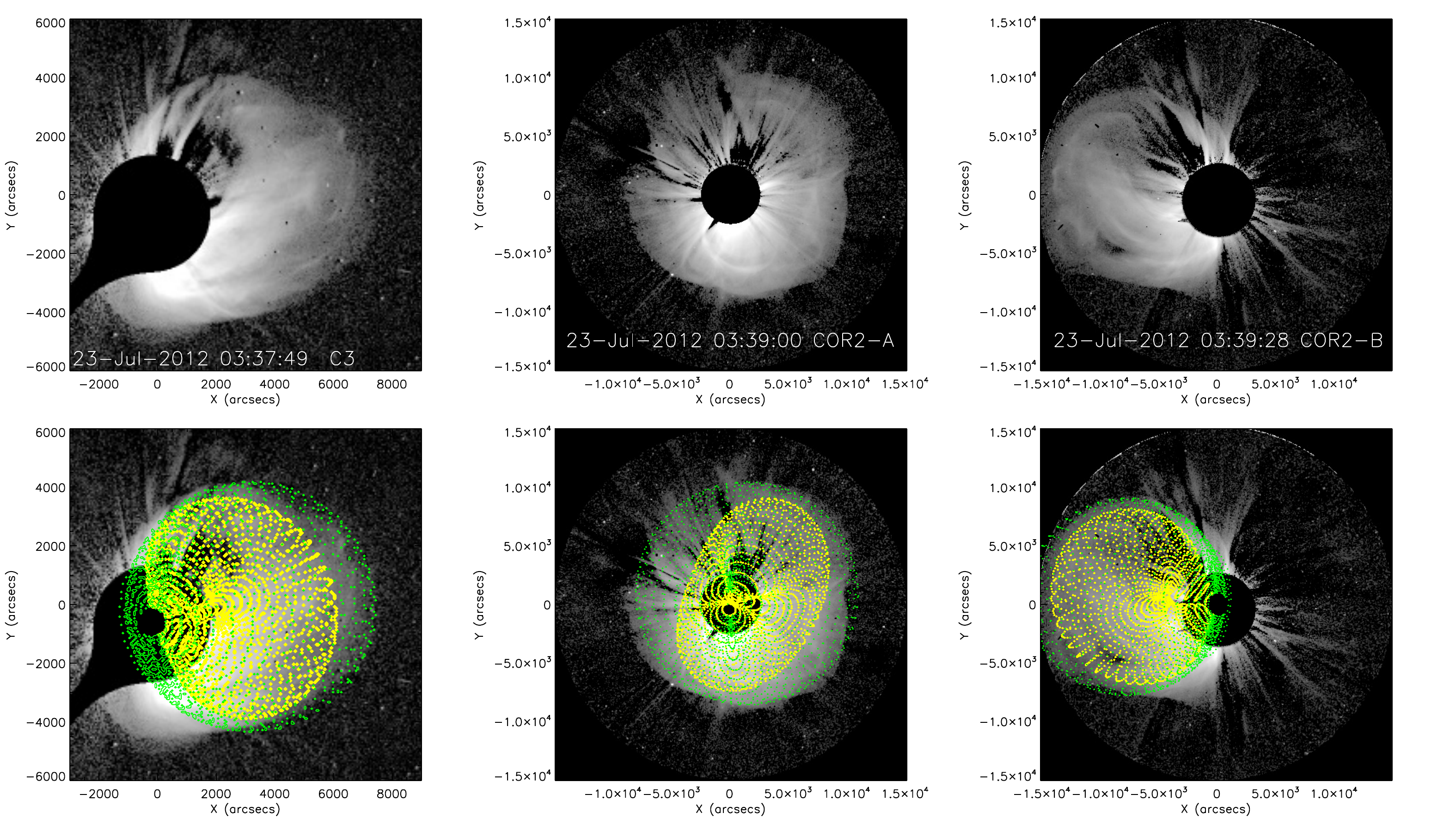}}
 \caption{Same as Figure~\ref{gcs} but for the image triplet COR2-A, COR2-B, and LASCO-C3 at $\sim$03:39~UT. }
	\label{gcs2}
\end{figure}

For investigating the propagation behavior of the CME in interplanetary space, and to relate the results from remote sensing and in-situ data, we use the DBM \citep{vrsnak07,vrsnak13}. Applying the DBM we assume that the CME propagation speed ($v_{\rm CME}$) at far distances from the Sun is governed only by the conditions of the ambient solar wind flow (density, $\rho_{\rm SW}$, and speed, $v_{\rm SW}$) which can be expressed in terms of the magnetohydrodynamical analogue of the aerodynamic drag \citep[for more details see][]{cargill96}. The acceleration due to the aerodynamic drag of the ambient solar wind is given by $a_{\rm D}$\,=\,$\gamma(v_{\rm CME}-v_{\rm SW})|v_{\rm CME}-v_{\rm SW}|$ with $\gamma$\,=\,$C_{\rm D}({A_{\rm CME}\rho_{\rm SW}}/{m_{\rm CME}})$ where $A_{\rm CME}$ and $m_{\rm CME}$ is the cross-section and the mass of the CME, respectively, $\rho_{\rm SW}$ the ambient solar wind density and $C_{\rm D}$ is the dimensionless drag coefficient, typically of order unity \citep[\textit{cf.}][]{cargill04}. The statistically derived $\gamma$ values for magnetic ejecta have a range of 0.1--2$\times$10$^{-7}$~km$^{-1}$ and $v_{\rm SW}$ of 400--500~km~s$^{-1}$ \citep{vrsnak13}. A recent statistical study shows that for the shock structures of CMEs the analytical DBM yields similar results as the numerical WSA-ENLIL+Cone model \citep[see \textit{e.g.},][]{odstrcil99} when using the parameter combination $\gamma$\,=\,0.1$\times$10$^{-7}$~km$^{-1}$ and $v_{\rm SW}$\,=\,400~km~s$^{-1}$ \citep[see][]{vrsnak14}. An online version of the DBM is available under \url{http://oh.geof.unizg.hr/DBM/dbm.php}.

\section{Results}\label{res}

Due to the close location of the source region of the CME (W141$^{\circ}$) and the plane-of-sky viewing angle of STEREO-B (W155$^{\circ}$), it is reasonable to assume that STEREO-B images are only marginally affected by projection effects. From high-cadence EUVI-B 171\AA~data (1.25~min resolution) we are able to track the onset of a loop formation at 2:02~UT that further develops into the COR1-B field-of-view (FoV) revealing at 2:13~UT the first signature of the CME. The CME can be tracked into the FoV of COR2-B and is also identified in STEREO-A white-light as well as SoHO/LASCO data enabling us to derive its 3D kinematics.

\begin{figure}
\centerline{\includegraphics[width=1\textwidth,clip=]{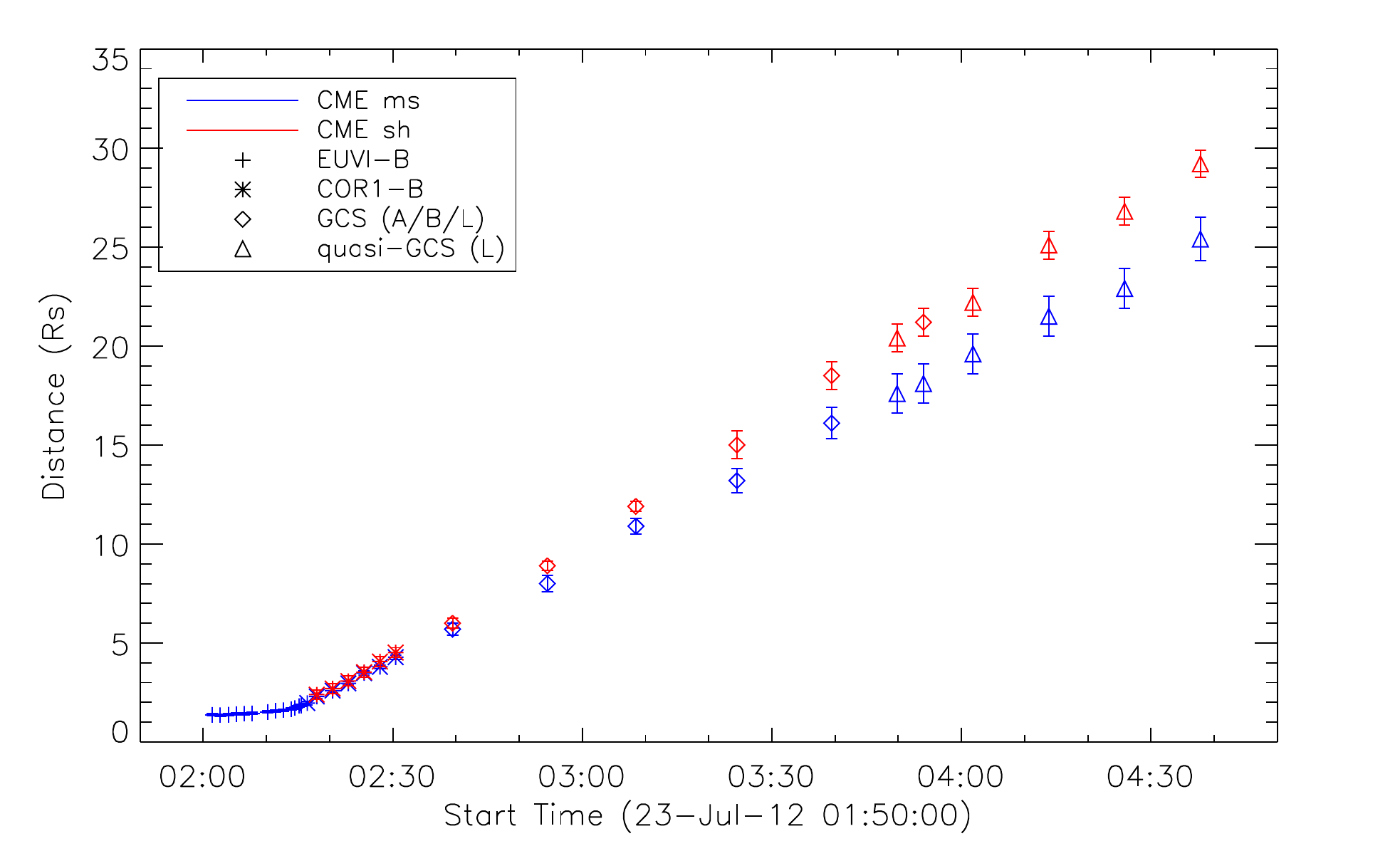}}
 \caption{3D distance-time measurements for the shock (red) and magnetic structure (blue) of the CME. The measurements for the CME were derived from different instruments (STEREO-A (A), STEREO-B (B), LASCO (L)) and reconstruction methods as given in the legend.  }
	\label{distance}
\end{figure}

The top panels of Figure~\ref{gcs} show coronagraph images from LASCO-C2, (left panel), COR2-A (middle panel), and COR2-B (right panel). The density envelope of the CME nicely reveals the shock-sheath and the magnetic structure (see annotation in the top right panel of Figure~\ref{gcs}). In order to compare the in-situ signatures of the arriving shock and magnetic structure separately, we make for each feature a separate GCS reconstruction. The forward fit shown in Figures~\ref{gcs} and~\ref{gcs2} gives the best match with the white-light data as viewed from different vantage points. From this we find that the CME propagates in longitude along W125--135$^{\circ}$ and in latitude along N00--N10$^{\circ}$. The flux rope geometry of the magnetic structure is characterized by its face-on and edge-on width for which we derive 130$\pm$5$^{\circ}$ and 60$\pm$5$^{\circ}$, respectively, and its tilt angle relative to the solar equator with 60$\pm$5$^{\circ}$. To track the kinematical evolution beyond the FoV of COR2, we assume self-similar expansion, i.e.\,keeping all model parameters constant with exception of the distance, and fit the obtained CME geometries (shock, magnetic structure) to LASCO-C3 image data (we refer to these measurements as quasi-GCS). Measurements of the early evolution phase of the CME from EUVI-B 171\AA~and COR1-B images are deprojected by using a multiplication factor of 1.1 to match the 3D values. With this we derive deprojected distance-time data from $\approx$1.25~R$_{\odot}$ up to 30~$R_{\odot}$. We intentionally used some extreme scaling to track the shock front of the CME, however, we note that a clear separation between shock and magnetic structure is only observed in COR2 and LASCO data.

\begin{figure}
\centerline{\includegraphics[width=1\textwidth,clip=]{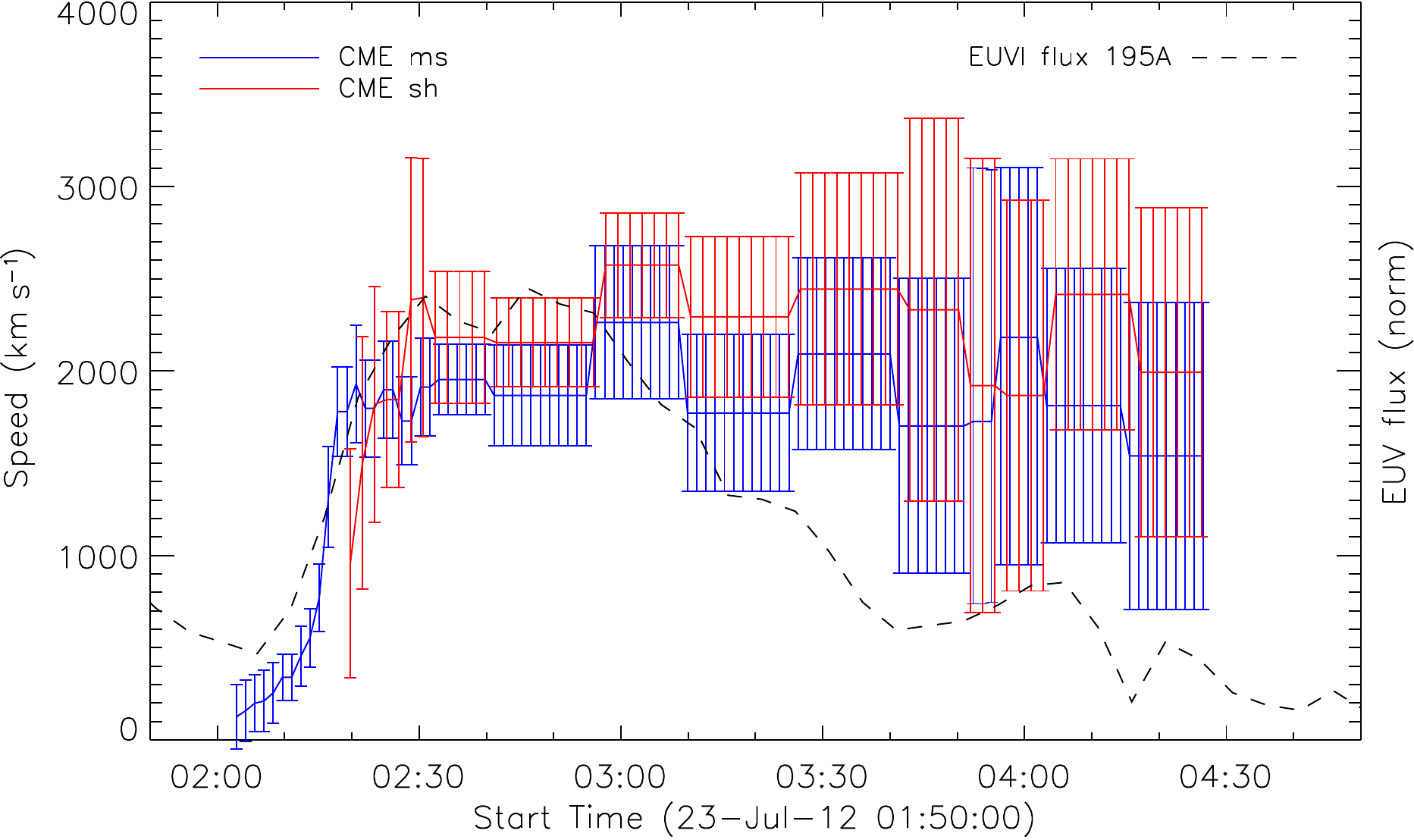}}
 \caption{3D speed-time profile derived from the distance-time data for the CME shock (sh) and magnetic structure (ms). The normalized EUVI-A 195\AA~flux is shown as black dashed line. }
	\label{speed}
\end{figure}

\begin{figure}
\centerline{\includegraphics[width=1\textwidth,clip=]{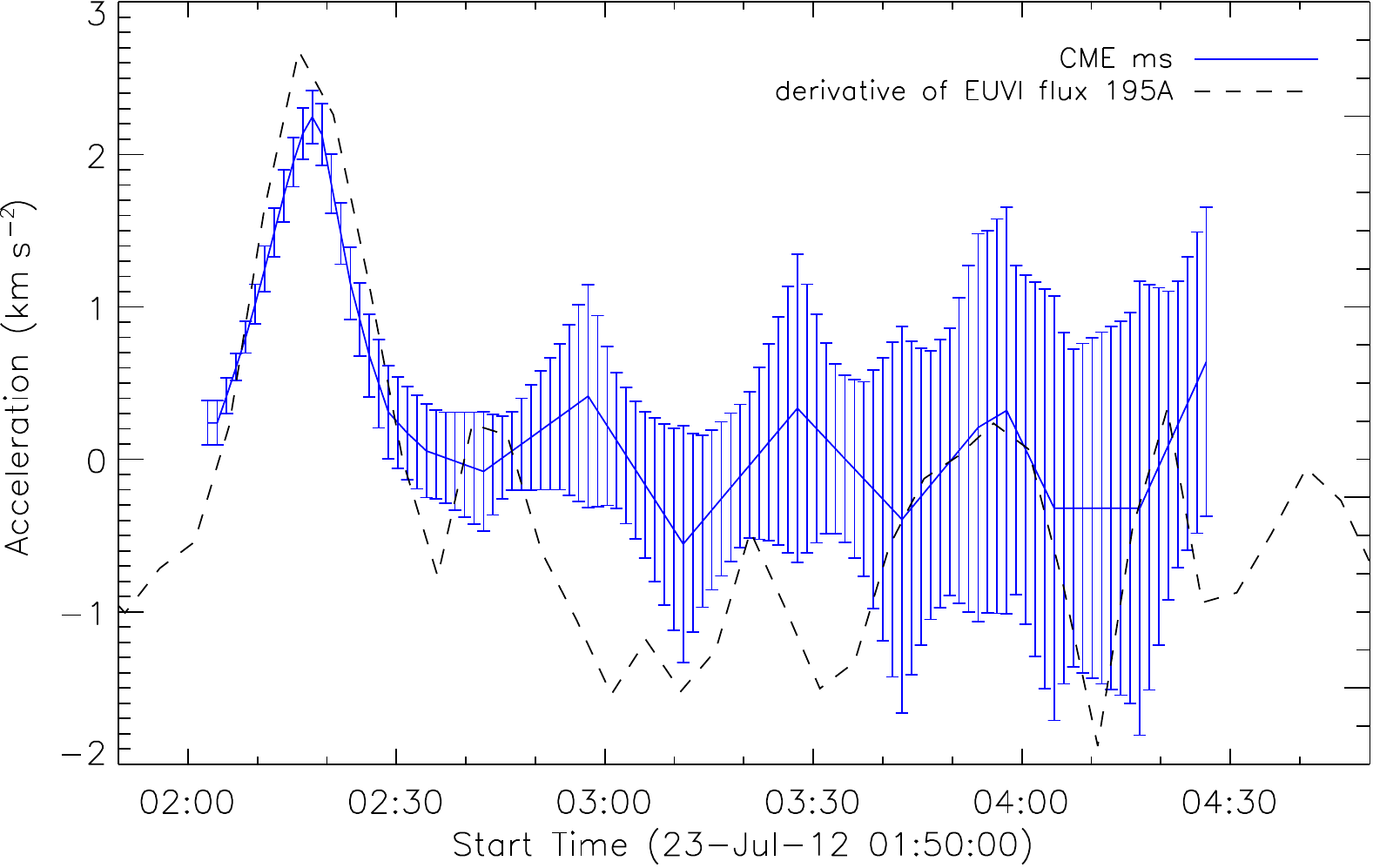}}
 \caption{3D acceleration-time profile derived from the distance-time data for the magnetic structure (ms) of the CME. The derivative of the EUVI-A 195\AA~flux (normalized) is shown as black dashed line.}
	\label{accel}
\end{figure}

Figure~\ref{distance} gives the deprojected distance-time values resulting from the GCS model and individual measurements (quasi-GCS and data from a single spacecraft). Due to the uncertainty in the 3D reconstruction, especially for distances far from the Sun, we put rather conservative error bars which is reflected in the uncertainties derived for the speed- and acceleration-time profiles. Figure~\ref{speed} shows the speed-time profile together with the EUV flux in the 195\AA~channel. The CME shock front reaches around 03:00~UT ($\pm$5~min) a maximum speed of 2580$\pm$280~km~s$^{-1}$ and the magnetic structure 2270$\pm$420~km~s$^{-1}$. Figure~\ref{accel} shows the acceleration-time profile of the magnetic structure together with the derivative of the EUV flux in the 195\AA~channel\footnote{As the shock and magnetic structure can be clearly distinguished starting from COR2 FoV, the shock acceleration profile is unreliable and not used for further analysis.}. The maximum acceleration yields 2.25$\pm$0.18~km~s$^{-2}$ with an acceleration duration of $\approx$30~min. Consistent with case and statistical studies, the hard X-ray as well as the derivative of the soft X-ray flux of the flare is closely related to the acceleration profile of the associated CME \citep{temmer08,temmer10,bein12}. This gives strong evidence that the complex CME eruption is launched during the initial rising phase of the EUV emission. The CME mass is calculated from COR2-B data at a distance of about 14~$R_{\odot}$ and is found to be in the range of 1.5$\pm$0.5$\times10^{16}$~g. The CME mass refers to a lower limit since we neglect possible projection effects \citep{vourlidas00}. From the CME mass and maximum speed we derive the kinetic energy which is of the order of $\approx$5$\times$10$^{32}$~ergs. These values are at the head of energy distributions found for CME observations over three decades \citep{vourlidas11} and demonstrate how exceptional the event under study is.

\begin{figure}
\centerline{\includegraphics[width=1\textwidth,clip=]{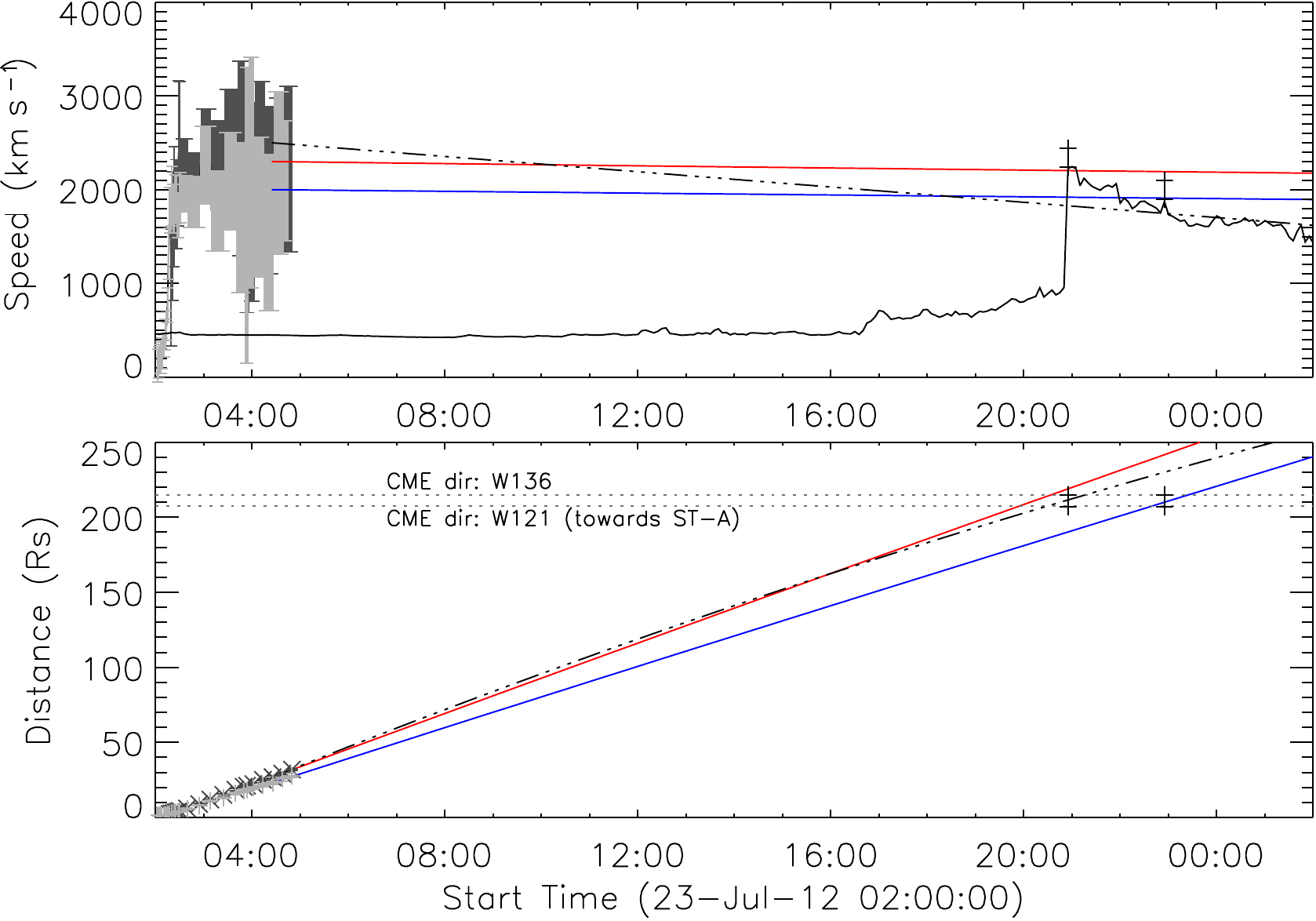}}
 \caption{Speed-time profile (top panel) and distance-time profile (bottom panel) for the observed CME shock (dark grey) and magnetic structure (light grey), DBM shock (red), and DBM magnetic structure (blue line). In the top panel the in-situ proton bulk speed (black line) is given together with the uncertainties for the impact speeds. Results from the empirical acceleration-velocity relation by \cite{gopal01} are marked as black dashed-dotted lines. The horizontal dashed lines in the bottom panel refer to the distance range that the CME needs to propagate owing to different propagation directions W121--W136$^{\circ}$.}
    \label{dbm_dir}
\end{figure}

 \begin{figure} 
\centerline{\includegraphics[width=1\textwidth,clip=]{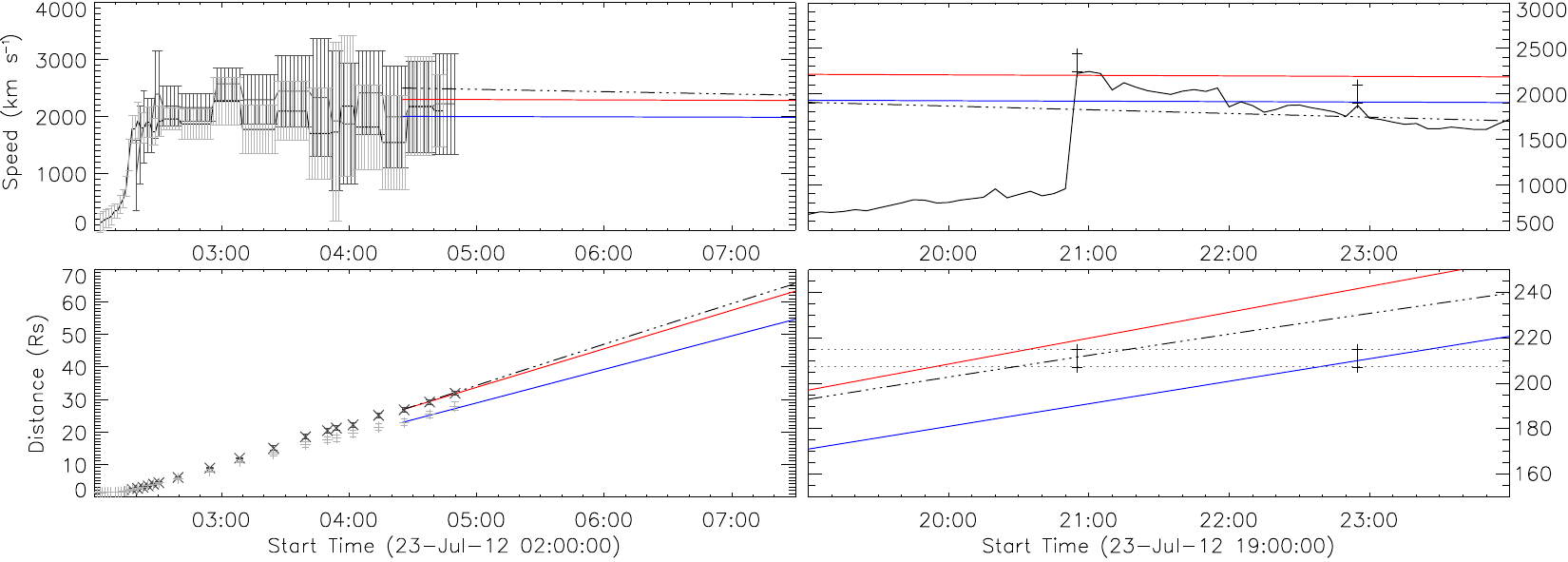}}
 \caption{Same as Figure~\ref{dbm_dir} but giving a close-up view on the early evolution (left panels) and the in-situ arrival (right panels).}
    \label{dbm-zoom}
\end{figure}

Figures~\ref{dbm_dir} and~\ref{dbm-zoom} show the results for the propagation behavior of the CME shock and magnetic structure using the DBM. The horizontal dashed lines in the bottom panel give the location of the STEREO-A spacecraft at W121$^{\circ}$ with a distance of 206~$R_{\odot}$ and the direction W136$^{\circ}$ with a CME propagation distance of 219~$R_{\odot}$, referring to that the apex of the CME is not directed towards STEREO-A \citep[see \textit{e.g.},][]{moestl13}. As comparison, we also show the result from the empirical acceleration-velocity relation, $a=2.193-(0.0054 \times v)$, proposed by \cite{gopal01}.

From observations we obtain that the CME shock reached at 23 July 2012 04:25~UT a distance of 27~$R_{\odot}$ with a speed in the range of 1100--2900~km~s$^{-1}$ (\textit{cf.}\,Figures~\ref{distance} and~\ref{speed}) and arrives at STEREO-A with a distance of 206~$R_{\odot}$ the same day at 20:55~UT having an impact speed of 2250~km~s$^{-1}$ \citep[\textit{cf.}][]{russell13}. The DBM input parameters are constrained by the remotely observed parameters and the model output is controlled by the in-situ observations. The model parameters which show the best match with the observations are summarized in Table~\ref{shockDBM}. From this it follows that the ambient solar wind flow speed in which the fast shock is propagating is of average value with 450~km~s$^{-1}$ close to the slow solar wind speed. The initial shock speed is required to be close to the impact speed (almost no deceleration) since for higher initial shock speeds, larger than $\approx$2300~km~s$^{-1}$, the model could not reproduce the observed arrival time and would arrive much too early. We need to apply a low $C_{\rm D}$ value of the order of 0.5--0.8 which refers to a very heavy and fast CME as also found from MHD simulations by \cite{cargill04}. A remarkable result is obtained for the $\gamma$ value with 0.01$\times$10$^{-7}$~km$^{-1}$ it is lower by one order of magnitude compared to the statistical results given in \cite{vrsnak14}. We conclude for the shock, the weak deceleration is owing to the extremely low $\gamma$ value, which we will examine in more detail below. We would like to add, the empirical acceleration-velocity relation by \cite{gopal01} would give a correct arrival time using as input speed 2500~km~s$^{-1}$ but would have largely underestimated the impact speed.

We assume that the magnetic structure, as tracked in our study, coincides with the magnetic cloud which reached STEREO-A at 22:55~UT having an impact speed of 1870~km~s$^{-1}$ \citep[\textit{cf.}][]{russell13}. We constrain the model input parameters by observations from remote sensing data which give for the magnetic structure a distance of 23~$R_{\odot}$ at 04:25~UT and a speed in the range of 750--2350~km~s$^{-1}$ (\textit{cf.}\,Figures~\ref{distance} and~\ref{speed}). As the magnetic structure propagates in the wake of the fast shock we use the resulting DBM speed-time profile from the CME shock as input for the ambient solar wind speed. This further constrains the model input values. We obtain the best match by using the DBM parameters as given in Table~\ref{msDBM}. The initial speed of the magnetic structure is required to be close to the observed in-situ impact speed ($\approx$2000~km~s$^{-1}$). Similar as derived for the CME shock, the deceleration of the magnetic structure is very small but required in order to match the correct arrival time as measured in-situ.  $C_{\rm D}$ is of the typical value of unity and $\gamma$ reveals to be of 0.15$\times$10$^{-7}$~km$^{-1}$, a reasonable value for massive CMEs. We conclude that for the magnetic structure, the weak deceleration is produced by the high ambient solar wind speed that actually resembles the CME shock speed.

\begin{table}
\caption{DBM and observational parameters for the shock of the CME. The parameters time ($t$), distance ($d$), and speed ($v$) as derived from observations constrain the DBM. Further DBM parameters are the ambient solar wind speed ($v_{\rm SW}$), the drag parameter ($\gamma$)  as well as the dimensionless drag coefficient ($C_{\rm D}$). The given parameters show the best match found between model and observations. DBM$_{\rm out1}$ refers to a best match between model and in-situ observations for the parameters distance and speed, DBM$_{\rm out2}$  for time and speed.
}
\label{shockDBM}
\begin{tabular}{ccccccc} 
  \hline
                  & $t$ [UT] & $d$ [Rs] & $v$ [km s$^{-1}$] & $v_{\rm SW}$ [km s$^{-1}$] &$\gamma$ [10$^{-7}$~km$^{-1}$] & $C_{\rm D}$ \\ 
  \hline
Remote obs & 04:25 &  27 & 1100--2900  & --- & --- & ---\\
DBM$_{\rm input}$ & 04:25   &  27 & 2300 & 450 & 0.01 & 0.5--0.8 \\
In-situ obs & 20:55  & 206 &  2250 & --- & --- & ---\\
DBM$_{\rm out1}$ & 19:45   &  206 &  2210 & --- & --- & ---\\ 
DBM$_{\rm out2}$ & 20:50   &  217 &  2200 & --- & --- & ---\\
  \hline
\end{tabular}
\end{table}

 \begin{table}
\caption{The same as Table~\ref{shockDBM} but for the magnetic structure of the CME. }
\label{msDBM}
\begin{tabular}{ccccccc} 
  \hline
& $t$ [UT] & $d$ [Rs] & $v$ [km s$^{-1}$] & $v_{\rm SW}$ [km s$^{-1}$] &$\gamma$ [10$^{-7}$~km$^{-1}$] & $C_{\rm D}$ \\ 
  \hline
Remote obs & 04:25 & 23 &  750--2350 & --- & --- & --- \\
DBM$_{\rm input}$ & 04:25 & 23 & 2000 & $v_{\rm DBM(shock)}$ & 0.15 & 1   \\
In-situ obs & 22:55 & 0.96 & 1870 & --- & --- & --- \\
DBM$_{\rm out1}$ & 22:30 &  206  & 1910 & --- & --- & --- \\ 
DBM$_{\rm out2}$ & 22:55 & 211 & 1910 & --- & --- & --- \\
  \hline
\end{tabular}
\end{table}

\begin{figure}
\centerline{\includegraphics[width=1\textwidth,clip=]{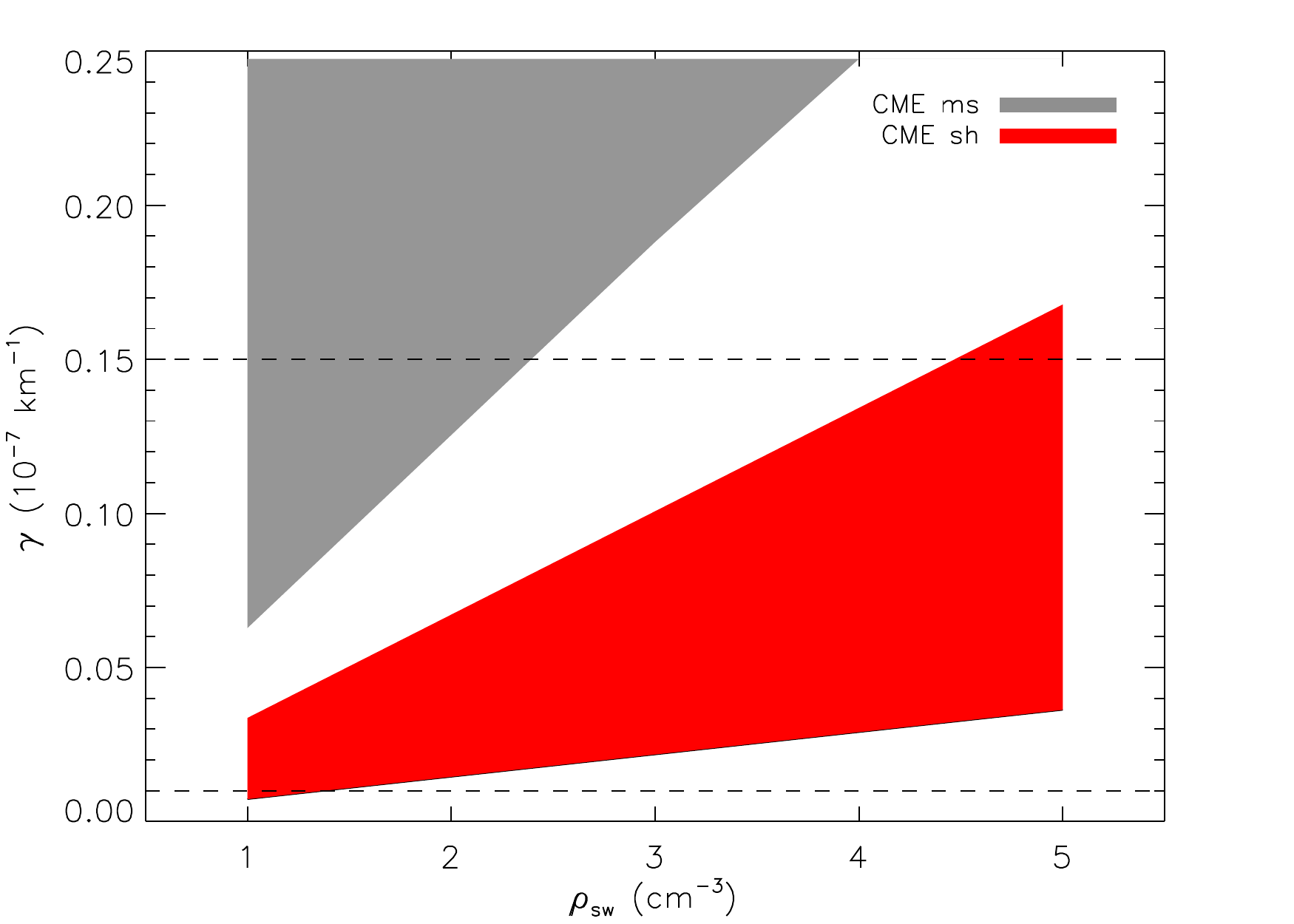}}
 \caption{Calculated $\gamma$ values for the CME shock (red area) and magnetic structure (grey area) versus different values for $\rho_{\rm sw}$. The dashed lines mark the $\gamma$ values as applied for the DBM. The given range for $\gamma$ covers uncertainties in the derived CME mass, cross-section area, and $C_{\rm D}$. Minimum values for $\gamma$ are obtained applying the upper limit in CME mass, the lower limit in CME size, and the lower limit for the $C_{\rm D}$. }
    \label{gamma}
\end{figure}

The question remains, is the extremely low $\gamma$ value as applied for the CME shock physically meaningful? We examine $\gamma$ for the CME shock and the magnetic structure by calculating it directly from the individual variables. We use the observed CME mass, the resulting CME geometry from the GCS model to obtain the cross-section, and the empirical relation by \cite{leblanc98} to derive the ambient solar wind density. For the CME shock the cross section ($A_{\rm sh}$) is defined as the plane area of a spherical cap with the stand-off distance between the shock and the magnetic structure as height of the cap. The stand-off distance $s$ is assumed to be linearly related with the propagated distance $d$. From observations we derive $s\approx4~R_{\odot}$ at $d\approx30~R_{\odot}$ (\textit{cf.}\,Figure~\ref{distance}) and we make a linear extrapolation for the distance range up to 1~AU. Using simple geometrical relations we find $A_{\rm sh}=(2ds-s^2)\pi$. For the magnetic structure the cross-section ($A_{\rm ms}$) is an ellipse with the semi-minor and semi-major axes defined by the obtained GCS edge-on and face-on width ($w_{\rm eo}, w_{\rm fo}$), expanding in a self-similar manner up to 1~AU. The area of this ellipse can be calculated with $A_{\rm ms}=\tan(\frac{w_{\rm eo}w_{\rm fo}}{2})d^2\pi$. For the dimensionless parameter $C_{\rm D}$ we use the model results which are 0.5--0.8 for the shock and 1 for the magnetic structure. In Figure~\ref{gamma} we plot the derived $\gamma$ values, including the derived uncertainties for all the variables, against the solar wind proton density values normalized for the distance at 1~AU. From this we derive that a density of $\rho_{\rm sw}$\,=\,1--2~cm$^{-3}$ indeed yields for the shock $\gamma$ as low as 0.01$\times$10$^{-7}$~km$^{-1}$. For the magnetic structure $\rho_{\rm sw}$\,=\,1--3~cm$^{-3}$  is required to derive $\gamma$ in the range of 0.15$\times$10$^{-7}$~km$^{-1}$.

\section{Discussion}\label{dis}

Using the DBM, we investigate the evolution and propagation from the Sun up to STEREO-A for the fast CME which occurred on 23 July 2012. We apply the GCS model assuming a spherical geometry for the shock as well as an idealized flux rope for the magnetic structure and study their 3D kinematics. From this we obtain for the maximum speed of the CME shock 2580$\pm$280~km~s$^{-1}$ and for its magnetic structure 2270$\pm$420~km~s$^{-1}$. We note that these values are lower than those derived by \cite{liu14} with $\approx$3050$\pm$200~km~s$^{-1}$. \cite{liu14} use a triangulation technique of time-elongation maps which are extracted from the ecliptic plane. Our study is based on GCS modeling which reproduces the entire density envelope of the CME along its propagation direction, hence, the results are not restricted to structures propagating along the ecliptic. We note that deviations from self-similar expansion and the idealized geometry assumption might be the caveat for the GCS results.

From the kinematics we derive the acceleration phase of the CME. Considering for the major acceleration peak of the CME an acceleration duration of $\approx$30~min, we estimate from different power-law relations\footnote{According to statistics, the CME acceleration duration and peak value of acceleration are closely related parameters which are inversely proportional. } maximum acceleration values in the range of 0.2--1.0~km~s$^{-2}$ \citep{zhang06,vrsnak07acc,bein11}. However, the derived acceleration value from observations is much higher (2.2$\pm$0.1~km~s$^{-2}$), showing the exceptional characteristics of this event. From this we may speculate that most probably it is the underlying magnetic reconnection process itself that is able to efficiently drive the CME to such high speeds at far distance from the Sun. Related to this we note that in general CMEs that reach high peak accelerations start at lower coronal heights \citep{bein11} and suggest that the low starting height of the CME might contribute to the strong magnetic field as measured in-situ.

Assuming that the further CME evolution is solely governed by the drag force owing to the ambient solar wind (speed and density), we use the DBM to simulate the short travel time from Sun to STEREO-A as well as the high impact speed. The CME shock propagation is successfully reproduced by using reasonable model input values supported by the observational results. We derive that the CME hardly decelerates in interplanetary space and find that it is not necessarily ultra-fast. The model required CME speed input values should not exceed 2300~km~s$^{-1}$ in order to reproduce the observed arrival time at STEREO-A. The ambient solar wind speed is found to be of average value close to the slow solar wind speed and might not play the key role for producing the short propagation time. The extreme low drag exerted on the CME is due to the applied $\gamma$ value which has to be one order of magnitude lower than statistically derived for DBM shocks \citep{vrsnak14}. The very high mass of the CME under study is one key requirement for a low $\gamma$ value and the low drag \citep[see also][]{vrsnak10}. But most important, the ambient solar wind density should not exceed 1--2~cm$^{-3}$ in order to derive the $\gamma$ value as low as required for our study. With this we support the interpretation by \cite{liu14} who pointed out the importance of the preconditioning of interplanetary space as consequence of the prior CME from July 19, 2012. As shown in \cite{liu14} the trailing part of the prior CME has a density as low as 1~cm$^{-3}$. We further conclude that a largely radial orientation of the interplanetary magnetic field, due to its stretching, may reduce the pile-up of solar wind and delay its replenishment. This may preserve the preferable conditions for a fast propagation of the CME launched 3.5~days later. In this respect we note that owing to the low solar activity of cycle 24, the solar wind density is generally low. For the propagation behavior of the CME magnetic structure we find that the weak drag exerted on the magnetic structure is due to the high ambient solar wind speed resembling its propagation in the wake of the fast shock.

\section{Conclusion}\label{con}

We identify the following key characteristics in the 23 July 2012 CME:
\begin{itemize}
\item high CME launch speed due to high flare energy release of long duration 
\item very high CME mass due to high amount of filament plasma 
\item extremely low ambient solar wind density due to previous CME and weak solar cycle
\end{itemize}

In conclusion, the extreme character of the 23 July 2012 eruption is first of all marked by the long duration flare energy release. Due to an efficient magnetic reconnection process the CME might reach a very high speed that is sustained by the prolonged energy release. The underlying mechanism which is able to build up the energy in the source region as well its conversion into such high kinetic energy is beyond the limit of information from observational data. We encourage modelers to provide further insight into the physics of source regions and reconnection processes related to such extreme events. The very high mass of the CME, related to the filament eruption, as well as the preconditioning of interplanetary space, in terms of reduced solar wind density, may have been the decisive factors for low deceleration leading to the short transit time and high in-situ impact speed. 


\begin{acks}
M.T. acknowledges the Austrian Science Fund (FWF): P20145-N16. N.V.N's work has been supported by NSF grant AGS-1259549, NASA AIA contract NNG04EA00C, and the NASA STEREO mission under NRL Contract No.\,N00173-02-C-2035. We appreciate the provision of PLASTIC data supported by NASA Grant NNX13AP52G. We thank Y.D.~Liu for valuable comments on the manuscript and J.G.\,Luhmann, as well as Y.\,Li, and B.J.\,Lynch for helpful discussions.
\end{acks}

   
\bibliographystyle{spr-mp-sola}
\tracingmacros=2

\end{article} 

\end{document}